\documentstyle[prl,aps,epsf]{revtex}

\begin{document}

\title{Coulomb effects in a ballistic one-channel S-S-S device}

\author{D. A. Ivanov$^{1,2}$,  M. V. Feigel'man$^1$ }

\address{ $^1$ L.D.Landau Institute for Theoretical Physics, 117940
Moscow, Russia \\
$^2$ 12-127 M.I.T. Cambridge, MA, 02139 USA}
\date{November 19, 1997}

\maketitle

\begin{abstract}

We develop a theory of Coulomb oscillations in superconducting devices
in the limit of small charging energy $E_C \ll \Delta$.
We consider a small superconducting
grain of finite capacity connected to two superconducting leads
by nearly ballistic single-channel quantum point contacts.
The temperature is supposed to be very low, so there are no
single-particle excitations on the grain.
Then the behavior of the system may be described as quantum
mechanics of the superconducting phase on the island.
The Josephson energy as a function of this phase has two minima which become
degenerate at the phase difference on the leads equal to $\pi$,
the tunneling amplitude between them being controlled by
the gate voltage at the grain.
We find the Josephson
current and its low-frequency fluctuations and predict their
periodic dependence on the induced charge $Q_x=C V_g$ with
period $2e$.

\end{abstract}

\section{Introduction.}

Coulomb effects in several different types of
 three-terminal devices consisting of an island connected
to external leads by two
weak-link contacts, and capacitatively coupled to an
 additional gate potential, have been extensively studied  during last
years.  The systems with a normal-metal island and leads 
 were studied theoretically both in the tunnel-junction limit \cite{MG1}
and in the case of a quantum point contact with almost perfect transmission 
\cite{M}.  The theory of charge-parity effects and Coulomb modulation of 
the Josephson current was investigated in details in \cite{GM}. All the
above-mentioned systems at present are realized experimentally.
Recently it was shown to be possible to produce quantum point contact
between two superconductors via a normal-conductive region made of
two-dimensional electron gas (2DEG) \cite{jap}; smeared step-wise
behaviour of the critical current was observed, in qualitative agreement
with predictions \cite{BH} for the superconductive quantum contact with
a few conduction channels of high transmittivity. 
An observation of a non-sinusoidal current-phase relation in 
superconducting mechanically-controllable break junctions has been reported
in \cite{K}, again in agreement with \cite{BH}.
Another interesting experimental achivement
was reported in \cite{kasumov}, where  S-N-S contact with a size comparable
to the de Broghle wavelength in the N region made of BiPb was realized
and nonmonotonic behaviour of the critical current with the thickness of
 normal region was found. This remarkable development of technology points
to the principal possibility to make a system of a small superconductive (SC)
island connected to the superconductive leads by  two quantum point contacts
(QPC).
In such a system macroscopic quantum effects due to competition between 
Josephson coupling energy and  Coulomb (charging) energy could be realized 
together with quantization (due to small number of conductive channels)
 of the Josephson critical current. 

In the present paper we develop
a theory for an extreme case of such a system, namely, for the case of
two almost ballistic one-channel QPCs connecting a small SC island
with two SC leads. We consider the limit of the characteristic
charging energy much smaller than the superconducting gap, $E_C\ll\Delta$,
and, therefore, the Coulomb effects are small.
We derive the dependences of the average Josephson current
across the sytem, and its fluctuations (noise power) as functions of the SC
 phase difference between the leads $\alpha$, and of the electric gate 
potential $V_g$. 
The Coulomb effects reveal themselves at phase differences 
$\alpha$ close to $\pi$, when the two lowest states are almost 
degenerate.
 We show that such a system realizes a tunable
quantum two-level system (pseudo-spin 1/2) which may be 
useful for the realization of quantum computers
(see e.g. \cite{kitaev,devinch,schoen,qaverin} ).

The paper is organized as follows. We start with considering a single
QPC connecting a superconducting island to a single lead (Section II). 
We find the oscillations of the effective capacitance on the island  as a 
function of the gate potential (in some analogy with Matveev's 
results \cite{M} for a normal QPC). Depending on the backscattering
probability in the contact, it may be described either in adiabatic
or in diabatic approximation. We find the condition for the
diabatic-adiabatic crossover. Then in Section III we formulate a 
simple model for the double-contact system in the adiabatic approximation.
We replace the full many-body problem by a quantum-mechanical problem 
for the dynamics of the SC phase  on the middle island.
In Sec.IV we calculate average Josephson current through the
 system as a function of $\alpha$ and $V_g$, with a particular emphasis
on the case of the phase difference $\alpha$ close to $\pi$ (when our 
effective two-level system is almost degenerate). Sec.V is 
devoted to the analysis of the Josephson current noise; we calculate
integrated intensity $S_0$ of the "zero"-frequency noise (an analogue of the
noise calculated in \cite{MR1,AI,Gordei} for a single superconduvtive QPC)
as well as finite-frequency noise $S_{\omega}$ due to transitions between the
 two almost-degenerate levels. Finally, we present our conclusions in
Sec.VI.

\section{Adiabatic-diabatic crossover in a single superconducting
quantum point contact.}

Consider a small superconducting island connected to an external
superconducting lead by an one-channel nearly ballistic quantum point
contact \cite{BH,B}.
The electric potential of the grain may be adjusted via
a gate terminal (fig.~1a).
Following \cite{BH} we assume that the contact
is much wider than the Fermi wavelength (so that the transport through
the constriction may be treated adiabatically), but much smaller than
the coherence length $\xi_0\equiv\hbar v_F/\pi\Delta$ (where $v_F$ is the
Fermi velocity, $\Delta$ is the superconducting gap).


\begin{figure}
\centerline{\epsffile{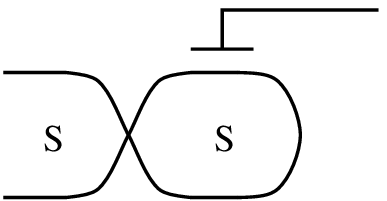}\hskip 3cm \epsffile{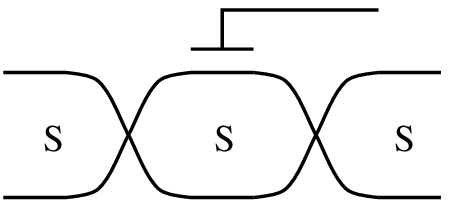}}
\caption{(a) Single QPC. The system consists of a SC grain
connected to a SC lead via a QPC. A gate terminal is used to control
the electric potential of the grain.
(b) Double-contact S-S-S system. The second terminal is added to the
single-QPC setup.}
\label{fig1}
\end{figure}

Our assumption of low temperature is that the average number of
one-electron excitations on the island is much less
than one. Then they cannot contribute to the total charge of the
grain and we may restrict our Coulomb blockade problem to the evolution
of the superconducting phase only. The condition of low temperature
is then $T < \Delta / \log (V \nu(0) \Delta)$, where
$V$ is the volume of the grain, $\nu(0)$ is the density of electron
states at the Fermi level.

We neglect phase fluctuations in the bulk of the island and
describe the whole island by a single superconducting phase $\chi$.
At a fixed value of the phase on the island, the spectrum of the
junction consists of the two Andreev states
localized on the junction and the continuum spectrum
above the gap $\Delta$ \cite{B} (fig.~2).
The energies of the Andreev states lie
below the gap:
\begin{equation}
E(\chi)=\pm\Delta\sqrt{1-t\sin^2(\chi/2)},
\label{Andreev}
\end{equation}
where $\chi$ is the phase difference at the contact,
$t$ is the transmission coefficient.


\begin{figure}
\centerline{\epsffile{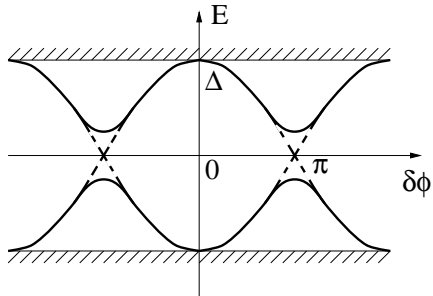}}
\caption{Single-contact energy spectrum.
The spectrum consists of the continuum of delocalized states
and the two Andreev (subgap) states. Dashed lines denote Andreev
states in the absence of backscattering (diabatic terms). Solid lines
are the states split by backscattering (adiabatic terms).}
\label{fig2}
\end{figure}

At $t=1$, the spectrum of Andreev states (\ref{Andreev}) 
has a level crossing point at $\chi=\pi$.
At this point, the left and right Andreev states have equal
energies, but in the absence of backscattering ($t=1$) the transitions
between them are impossible. Therefore, we expect that an ideal ballistic
contact cannot adiabatically follow the ground state
as the phase $\chi$ changes,
but remains on the same left or right Andreev state as it passes
the level-crossing point $\chi=\pi$.
We borrow the terminology from the theory of
atomic collisions \cite{lichten} and call the 
(crossing) Andreev levels at $t=1$ 
diabatic terms (dashed lines in fig.~2), and the split
levels --- adiabatic terms (solid lines in fig.~2).
Instead of transmission coefficient $t$, it will be more convenient
to speak of the reflection coefficient $r=1-t$. At $r=0$, 
the contact is described by diabatic terms. As $r$ increases,
the transitions occur between the terms, and at sufficiently large
$r$ the system will mostly adiabatically follow the split
Andreev levels. In this section we study adiabatic-diabatic
crossover and find the crossover scale for the reflection
coefficient $r$.

We assume that the reflection probability $r\ll 1$
(almost unity transmission) and that the charging energy $E_C\ll \Delta $
(the charging energy is defined by $E_C=(2e)^2/C$).
The latter assumption appears natural, because, like in the tunnel 
junctions \cite{LO} we expect
that the capacitance $C$ of the grain has an additional contribution from the
capacitance of the point contact. This capacitance is of order
$\Delta/e^2$. A more detailed discussion of this phenomenon will be
given elsewhere. At the moment we just mention that this contribution
to the capacitance leads to the inequality $E_C \leq \Delta$.

To probe the degree of adiabaticity, we study the periodic
dependence of the ground state energy $E_0$ on the gate voltage.
Because of the weakness of charging effects, this dependence will
be sinusoidal:
\begin{equation}
E_0(V_g)=\varepsilon \cos(2\pi N)     
\label{sinus1}
\end{equation}
(where $N=V_g C / 2e$ is the dimensionless voltage),
and we are interested in
the amplitude $\varepsilon$
of these oscillations. The physical meaning of this
periodicity is the oscillations of the induced charge on the
grain --- it follows immediately from the relation
\begin{equation}
\delta Q = {C\over 2e} {\partial E_0\over \partial N}.
\end{equation}

There is a simple physical explanation of the sinusoidal
dependence (\ref{sinus1}). The ground-state energy modulation is
determined by phase-slip processes in the contact. Such processes
are phase tunneling events with phase changing  by $\pm 2\pi$.
While the magnitudes of the clockwise and counter-clockwise
tunneling amplitudes are the same, their phases
are $\pm 2\pi N$. This results in the expression (\ref{sinus1}). 
Higher-order tunneling processes would give rise to higher-order harmonics
in the periodic $N$-dependence. This argument shows that the
amplitude of oscillations $\varepsilon$ coincides with the
phase-tunneling amplitude and, therefore, provides
a good measure of adiabaticity in the phase dynamics.

Under assumption $E_C\ll\Delta$, we may describe
the contact by the dynamics of the phase on the grain and thus
reduce the problem to a single-particle quantum mechanics.
Since we restrict our attention to low
lying excitations, it is only necessary to include the two Andreev
levels on the junction. 
The potential term is the Josephson energy of the Andreev
levels, the kinetic term is the charging energy. 
After a simple computation of the backscattering matrix
elements (the off-diagonal entries in the potential term), 
we arrive to the following Hamiltonian:
\begin{equation}
H=H(\chi)+{1\over 2}E_C(\pi_\chi - N)^2
\label{Ham3}
\end{equation}
where
\begin{equation}
H(\chi)=\Delta
\pmatrix{-\cos{\chi\over2} & r^{1/2}\sin{\chi\over2} \cr
r^{1/2}\sin{\chi\over2} & \cos{\chi\over2} }.
\label{Ham4}
\end{equation}
Here $\chi$ is the phase difference across the contact,
$r$ is the reflection coefficient. Obviously, the eigenvalues
of $H(\chi)$ reproduce the result (\ref{Andreev}).
The number of Cooper pairs at the grain $\pi_\chi$ is the momentum
conjugate to $\chi$, $[\chi,\pi_\chi]=i$. Notice that $\chi$ takes
values on the circle $\chi=\chi+2\pi$, and, accordingly, $\pi_\chi$
is quantized to take integer values. We may also write
$\pi_\chi=-i\partial/\partial\chi$.


This Hamiltonian loses its validity at the top of the upper band at
$\chi=2\pi n$, where the upper Andreev state mixes with the continuous
spectrum (fig.~2). Howerver, the probability of the phase $\chi$
to reach the top of the upper band of $H(\chi)$ is exponentially
small at $E_C\ll \Delta$ (smaller than the tunneling probability).
The adiabatic-diabatic crossover is determined by the properties
of the system near the minimal-gap point $\chi=\pi$. Therefore,
we may neglect the transitions to continuous spectrum
at $\chi=2\pi n$. 
At the same time, we must disregard tunneling porcesses
via the top of the upper Andrees band (next-nearest-neighbor
tunneling) which is present in the Hamiltonian (\ref{Ham3})-(\ref{Ham4}),
but not in the original system. The nearest-neighbor tunneling
is a feature of our model and is beyond the precision of our
approximation.

There are two opposite limits of the problem:
small and ``large'' reflection.

At zero reflection, the Hamiltonian splits into
lower and upper components. Within each component the potential
is periodic with the period $4\pi$.
As explained above, we must neglect the next-nearest-neighbor
tunneling via the top of the bands. Therefore, the potential
minima of $H(\chi)$ are disconnected and
cannot tunnel to each other, $\varepsilon=0$.


The opposite limit is the case of ``large'' reflection (the
precise meaning of "large reflection" consistent with $r\ll 1$ will be
clarified below). In this limit, the gap opens in the spectrum
of Andreev states, and the system adiabatically follows the
lower state. We can replace the two-level Hamiltonian $H(\chi)$
by its lowest eigenvalue and arrive to the quantum-mechanical problem of
a particle in a periodic potential. The quasiclassical limit of
this problem is solved in the textbook \cite{LL}.
In our notation the answer reads as follows:
\begin{equation}
\varepsilon_{ad}={\rm const} \sqrt{E_C \Delta}
\exp(-S_{cl}),
\end{equation}
where
\begin{equation}
S_{cl}=B_1 \sqrt{\Delta\over E_C}-{1\over 4}\log {\Delta\over E_C}
+ O(1)
\end{equation}
is the classical action connecting two nearest
minima (or more precisely the two return points).
The numerical constant $B_1$ is of order one
(at $r\to 0$, $B_1=4.69 + 1.41 r\log r +\dots $).

To study how the adiabaticity is destroyed it is useful to introduce
the dimensionless ``coherence factor'' $f(r)$ defined by
\begin{equation}
\varepsilon=f(r) \varepsilon_{ad},
\end{equation}
where $\varepsilon_{ad}$ is the amplitude of oscillations
of the ground-state energy
derived in the adiabatic approximation (with only the lowest Andreev
state included).
We see that $f(0)=0$, $f(r\gg r_{ad})=1$.
The crossover scale $r_{ad}$ can be derived by computing
the corrections to $f(r)$ in these two limits.

First consider the limit of weak backscattering ($r\ll r_{ad}$).
In this limit we take the wavefunction to be the ground state
of the Hamiltonian with zero $r$ (at a given wavevector $N$),
and then compute the first-order correction in $r^{1/2}$ to the
energy. The wavefunction is of ``tight-binding'' type and is
generated by the ``ground-state'' wavefunctions $\Psi_i$ localized
in the potential minima (diabatic terms). The components of the
two-dimensional vectors $\Psi_i$ alternate:
\begin{equation}
\Psi_i=\pmatrix{\Psi_i(\chi) \cr 0},
\qquad
\Psi_{i+1}=\pmatrix{0 \cr \Psi_{i+1}(\chi)}.
\end{equation}
Then we find
\begin{equation}
\varepsilon = 2 \langle \Psi_i | H_{12}(\chi) | \Psi_{i+1} \rangle
= 2 r^{1/2} \Delta \int d\chi \,
\Psi_i^*(\chi) \Psi_{i+1}(\chi) \sin{\chi\over2}
\label{diabatics}
\end{equation}
(We assume the wavefunctions $\Psi_i$ to be normalized).
It is important that
$\Psi_i$ and $\Psi_{i+1}$ are wavefunctions for different potentials
($-\Delta_0\cos(\chi/2)$ and $\Delta_0\cos(\chi/2)$) and
the overlap integral (\ref{diabatics})
has a saddle point at the minimal-gap point
$\chi=\pi$, and it reduces the effective region of integration
to $|\chi-\pi| \le (E_C/\Delta)^{1/4}$.
The normalization of the quasiclassical tail of the
wavefunctions $\Psi_i(\chi)$ yields
\begin{equation}
\Psi(\chi=\pi)= \exp(-S_{cl}(\chi=\pi)) 
\end{equation}
(up to a numerical factor independent of $E_C/\Delta$).
Thus we obtain
\begin{equation}
\varepsilon\sim r^{1/2} \Delta \left({E_C\over\Delta}\right)^{1/4}
\exp(-S_{cl}),
\end{equation}
i.e., in terms of the ``coherence factor'' $f(r)$,
\begin{equation}
f(r) \sim r^{1/2} \left({\Delta\over E_C}\right)^{1/4}. 
\end{equation}
The physical meaning of the integral (\ref{diabatics})
is the summation over all paths shown in fig.~3a.


\begin{figure}
\centerline{\epsffile{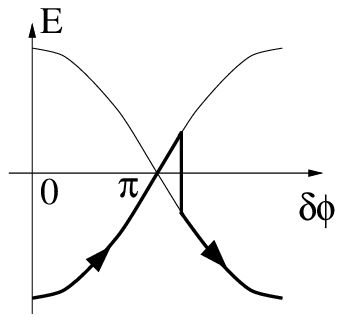}\hskip 3cm \epsffile{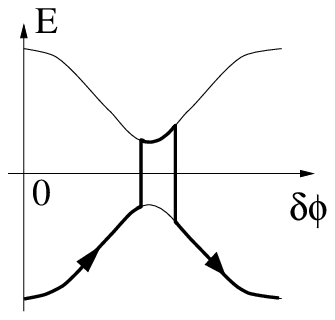}}
\caption{Tunneling paths in the diabatic (a) and
adiabatic (b) limits.
These diagrams represent the lowest-order corrections to the phase
tunneling amplitudes in the diabatic and adiabatic limits respectively.}
\label{fig3}
\end{figure}

The above calculation shows that the crossover scale to
adiabatic behavior is
\begin{equation}
r_{ad} \sim \left({E_C\over\Delta}\right)^{1/2}.   
\label{Rad}
\end{equation}

In fact, we neglected the effect of change
in the classical action $S_{cl}$
due to opening a gap; this effect
is estimated to be of order
\begin{equation}
\delta S_{cl} \sim  \sqrt{\Delta\over E_C} r \log r,
\end{equation}
i. e. it is a higher-order effect than
the change in $f(r)$ proportional to $r^{1/2}$.
Notice that the characteristic scale of this change in the
classical action is again $r_{ad}\sim \sqrt{E_C/\Delta}$
(corresponding to $\delta S_{cl}\sim 1$).

We may alternatively find the crossover scale $r_{ad}$
by computing the lowest order correction to the
``coherence factor'' $f(r)$ in the adiabatic limit.
In this limit the Hamiltonian (\ref{Ham3},\ref{Ham4}) may
be rewritten in adiabatic terms (the voltage $N$ is
for simplicity moved to the boundary condition
$\Psi(\chi+2\pi)=e^{2i\pi N} \Psi(\chi)$
by a gauge transformation)
as
\begin{equation}
H=-{E_C\over 2}({\partial\over\partial\chi})^2
+D(\chi) -{E_C\over2}[G(\chi){\partial\over\partial\chi}
+ {\partial\over\partial\chi} G(\chi)] -{E_C\over2}
G^2(\chi),
\label{Ham5}
\end{equation}
where
\begin{equation}
D(\chi)=\pmatrix{E_1(\chi) & 0 \cr 0 & E_2(\chi)}
\end{equation}
is the diagonalized form of the matrix (\ref{Ham4}), and
\begin{equation}
G(\chi)=\pmatrix{0 & g(\chi) \cr -g(\chi) & 0}, \quad
g(\chi)=\langle 0 | {\partial\over\partial\chi} | 1 \rangle,
\end{equation}
and $| 0\rangle$ and $|1\rangle$ are the eigenvectors of the
matrix (\ref{Ham4}).
The last term in the Hamiltonian (\ref{Ham5}) can be shown to give
smaller corrections than the term of the first order in $G(\chi)$.
A careful perturbation theory in $g(\chi)$ gives in second order
\begin{equation}
1-f(r) \sim \int_{\chi_1<\chi_2} e^{S_1(\chi_1, \chi_2) -
S_2(\chi_1, \chi_2)} g(\chi_1) g(\chi_2) \,
d\chi_1\, d\chi_2,
\label{integral1}
\end{equation}
where $S_{1,2} (\chi_1, \chi_2)$ are the classical actions
along the lower and the upper adiabatic branches between the
points $\chi_1$ and $\chi_2$.
This integral corresponds to summation over all tunneling paths
shown in fig.~3b.
The function $g(\chi)$ for
the given matrix $H(\chi)$ is a lorentzian peak at $\chi=\pi$
of height $r^{-1/2}$ and width $r^{1/2}$. Putting everything
together, the integral (\ref{integral1}) is calculated to
be
\begin{equation}
1-f(r) \sim {1\over r} \sqrt{E_C\over\Delta}.
\end{equation}
This asymptotics agrees with the found previously crossover
scale (\ref{Rad}).


To summarize the results of this section, the characteristic
scale for adiabatic-diabatic crossover in a nearly-ballistic
single contact is found to be $r_{ad}\sim \sqrt{E_C/\Delta}$.
The phase tunneling amplitude is proportional to the gate-voltage
 modulation of the effective capacitance of the island, and thus
can be directly measured. 
At low reflection
coefficients, these oscillations are proportional to
$\sqrt{r}$, like in the normal 1-channel QPC \cite{M}.

\section{Adiabatic approximation of a double-junction system.}

Now turn to the case of a double-junction system (fig.~1b).
As before, we assume that the reflection probabilities in both contacts
are small, $r_i\ll 1$, 
that the charging energy $E_C\ll \Delta $
and that the temperature is sufficiently low to prohibit
single-electron excitations on the grain. To adjust electrostatic
potential of the grain we again use a gate terminal, $N=V_g C/2e$
denotes the dimensionless gate voltage, as before.

For the moment, to simplify the discussion we assume that the reflection 
coefficients in the contacts are greater than the 
crossover scale $r_{ad}$ found in the previous section and, therefore,
we may consider only the lower adiabatic branch of the Andreev states.
In fact, the results may be extended further to the case $r_i<r_{ad}$
by using appropriate ``coherence factors'' $f(r)$, similar 
to those in the previous section.

We set the superconducting phase on one of the leads to be zero;
the phase on the other lead $\alpha$ is assumed to be fixed externally.
Then the total Josephson energy of the two contacts is (fig.~4):
\begin{equation}
U(\chi)=U_1(\chi)+U_2(\alpha-\chi),
\label{potential}
\end{equation}
where
\begin{equation}
U_i(\delta\phi)=-\Delta\sqrt{1-t_i\sin^2(\delta\phi/2)}
\label{groundstates}
\end{equation}
are the lower adiabatic Andreev terms in the two junctions.


\begin{figure}
\centerline{\epsffile{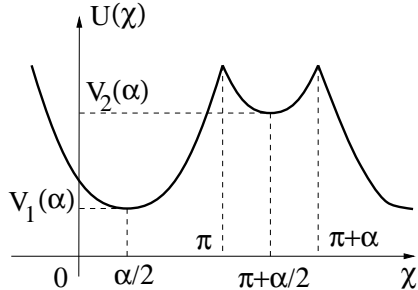}}
\caption{Potential $U(\chi)$. At $\alpha\ne0$ it has two minima. 
Finite backscattering in the contacts smoothes the summits of the
potential, but leaves the bottom of the wells unchanged.
}
\label{fig4}
\end{figure}

At $t_1=t_2=1$, the potential $U(\chi)$ obviously has two minima ---
at $\chi=\alpha/2$ and at $\chi=\alpha/2+\pi$ ---
and sharp summits at $\chi=\pi$ and $\chi=\pi+\alpha$ (fig.~4).
At small nonzero $r_i$, gaps open at the crossing points of Andreev
levels, which smoothes the summits of $U(\chi)$. Still, the bottom
of the potential remains practically unchanged.

The adiabatic Hamiltonian for the double
junction looks like follows:
\begin{equation}
H(\alpha,N)=U(\chi)+U(\alpha-\chi)+
{1\over 2}E_C(-i{\partial\over\partial\chi}-N)^2.
\label{Ham1}
\end{equation}
The potential term of the Hamiltonian is the sum of Josephson energies
of the contacts, the kinetic term is the Coulomb energy of the charge
at the grain.

\section{Josephson current.}

The condition
$E_C\ll\Delta$ allows us to treat the Coulomb term in the Hamiltonian
perturbatively.
First, neglecting the Coulomb term, we obtain a classical
system on the circle in the potential (\ref{potential}) with
two minima. 
The energies of the minima are
$V_1(\alpha)=-2\Delta |\cos(\alpha/4)|$ and $V_2(\alpha)=
-2\Delta |\sin(\alpha/4)|$ (see fig.~4). 
To a very good precision, we may neglect backscattering in determining
the minima --- except near the point $\alpha=0$. Since all the Coulomb 
effects occur near the resonance point $\alpha=\pi$, this approximation
is justified.
At zero temperature, our
classical system prefers
the lowest of the minima. Thus the energy of the S-S-S system in the
absence of the Coulomb term is given by
\begin{equation}
E(\alpha)=-2\Delta \cos(\alpha/4) \qquad
{\rm for} \quad -\pi<\alpha<\pi
\end{equation}
(see fig.~5). Differentiating this energy with respect to the phase
$\alpha$ gives the Josephson current
\begin{equation}
I(\alpha)=2e{\partial E(\alpha) \over \partial\alpha}=
\Delta \sin {\alpha\over 4} \qquad
{\rm for} \quad -\pi<\alpha<\pi
\end{equation}
(fig.~6). Notice that the current has large jumps at the points of
level crossing $\alpha=\pi+2\pi n$. Qualitatively this picture is
very similar to the case of a single S-S ballistic junction,
but the shape of the current-phase dependence $I(\alpha)$
is different.


\begin{figure}
\centerline{\epsffile{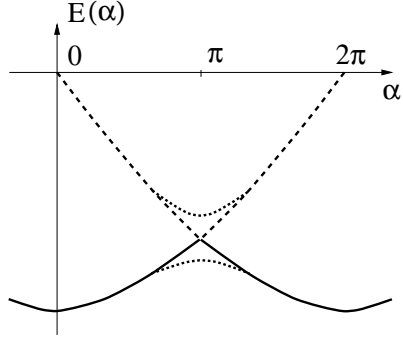}}
\caption{Classical minimum of the potential $U(\chi)$ as a function of the
external phase difference $\alpha$.
Dotted line shows the quantum gap opened by the Coulomb term.}
\label{fig5}
\end{figure}

\begin{figure}
\centerline{\epsffile{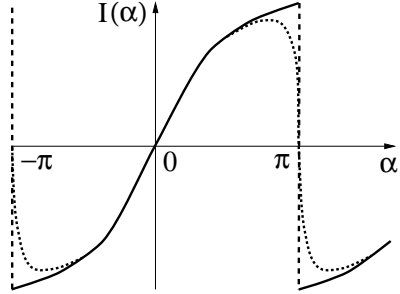}}
\caption{Josephson current as a function of the external phase difference 
$\alpha$.
Dotted line shows smearing of the singularity
due to the Coulomb term.}
\label{fig6}
\end{figure}

If we assume a non-zero temperature $T\ll\Delta$, the occupation of the
upper minimum is exponentially small except in the vicinity
of the level-crossing point $|\alpha-\pi|\sim T/\Delta$.
Thus, the effect of the temperature results in the smearing of
the singularity in $I(\alpha)$ at $\alpha=\pi$.

Another source of level mixing near the singular point $\alpha=\pi$
is {\it quantum} fluctuations, i.e. the fluctuations arising
from the kinetic term in the Hamiltonian (\ref{Ham1}). 
They result in nonzero amplitudes of tunneling through the two
potential barriers between the potential minima. Due to the shift
in the "angular momentum" by $N$, the wave functions in the two potential
wells aquire an additional factor $\exp(iN\chi)$. This results
in the relative phase of the two tunneling
amplitudes by $2\pi N$. The net tunneling amplitude
(defining the level splitting) may be written as
\begin{equation}
H_{12}(N)\equiv\Delta\gamma(N)=\Delta (\gamma_1 e^{i\pi N}
+ \gamma_2^{-i\pi N}).
\label{H12}
\end{equation}
where $\gamma_1$ and $\gamma_2$ are the two amplitudes
of phase tunneling in the two different directions
(i.e. of phase slip processes in the two different contacts).
Below we assume that these amplitudes are computed at the
level-crossing point $\alpha=\pi$, where they are responsible
for level splitting.

The amplitudes $\gamma_1$ and $\gamma_2$ obey all asymptotics
derived in the previous section (except for numerical
factors). When the backscattering in the contacts $r\gg r_{ad}$,
they may be found in the quasiclassical approximation:
\begin{equation}
\gamma_{1,2}\sim\left({E_C\over\Delta}\right)^{1/4}
\exp(-B_2\sqrt{\Delta\over E_C})\ll 1,
\end{equation}
where $B_2\sim 1$ is determined by the classical action connecting
the two potential minima (at $r \ll 1$, 
$B_2 \cong 1.45 + 2.20 r\log r + \dots$).
At $r\ll r_{ad}$, the tunneling amplitudes are
\begin{equation}
\gamma_{1,2}\sim r^{1/2}
\exp(-B_2\sqrt{\Delta\over E_C})
\label{gamma2}
\end{equation}

For the best observation of Coulomb oscillations, $\gamma_1$ and
$\gamma_2$ must be of the same order, but not very small.
In the ideal case $\gamma_1=\gamma_2=\gamma$ the total amplitude
\begin{equation}
\gamma(N)=2\gamma\cos(\pi N)
\label{gammas}
\end{equation}
Although the periodic dependence (\ref{gammas}) has
$4e$ period as function of the "external charge" $Q_x = CV_g \equiv 2eN$,
the Josephson current and its fluctuations depend on 
$|\gamma(N)|^2$ only (cf. eqs.(\ref{I},\ref{S}) below), 
and their period is $2e$ as expected~\cite{GM}.

The characteristic scale for the $r$-dependence
of $B_2$ is $\delta r \sim \sqrt{E_C/\Delta}$, therefore
for $\gamma_1$ and $\gamma_2$ to be of the same order,
the transparencies of the two contacts must differ by no more
than $|r_1-r_2|\le \sqrt{E_C/\Delta}$.

Here we should comment on the difference of our result 
(\ref{H12})-(\ref{gamma2}) from the normal two-channel
system discussed in \cite{M}. In the normal
system the two tunneling amplitudes multiply, and the net
ground-state energy oscillations are proportional to $r\ln{r}$
at small $r$. In the superconducting system, the external leads
have different superconducting phases, and the tunneling
in the two contacts occurs at different values of the phase
on the grain. Therefore, the tunneling amplitudes add with
some phase factors and give the asymptotic of $\sqrt{r}$ at $r\to 0$.
In fact, the oscillations in the superconducting system will
be proportional to $r$ (similarly to the normal system \cite{M}) in
a different limit --- at the phase difference $\alpha=0$,
when the potential $U(\chi)$ has a single minimum and a single barrier.

The hybridized energy levels in the vicinity of $\alpha=\pi$
are given by the eigenvalues of the $2\times 2$
Hamiltonian
\begin{equation}
H(\alpha,N)=\pmatrix{ V_1(\alpha) & H_{12} (N) \cr
                         H_{12} (N)  & V_2(\alpha) \cr}.
\label{Ham2}
\end{equation}
Diagonalization gives the two energy levels:
\begin{equation}
E_{1,2}(\alpha,N)=-\Delta\left[ |\sin{\alpha\over 4}| +
|\cos{\alpha \over 4}| \pm \sqrt{\left( |\sin{\alpha\over 4}| -
|\cos{\alpha \over 4}| \right)^2 + \gamma^2 (N)} \right],
\label{energies}
\end{equation}
the off-diagonal matrix elements of the Hamiltonian open a gap
at the level-crossing point $\alpha=\pi$ (fig.~5). 
This gap periodically
depends on the gate voltage $V_g$, and these oscillations comprise
the Coulomb effects in the S-S-S junction.

We can obtain the Josephson current by
differentiating the energy levels with respect to the phase $\alpha$.
The gap results in smearing the singularity in $I(\alpha)$ even at
zero temperature (fig.~6):
\begin{equation}
I(\alpha)=
{\Delta\over \sqrt2} \sin({\alpha-\pi\over 4})
\left[1-{\cos({\alpha-\pi\over 4}) \over
\sqrt{\sin^2({\alpha-\pi\over 4})
+ {1\over 2}\gamma^2(N)}} \right]
\qquad {\rm for} \quad \alpha\sim\pi.
\label{I}
\end{equation}
The width of the crossover at $\alpha=\pi$ depends periodically on $V_g$:
$|\alpha-\pi| \sim |\gamma(N)|$.

In the above discussion we neglected the excited oscillator states.
The interlevel spacing for the excitations in the potential
wells is of order $\sqrt{\Delta E_C}\gg\Delta\gamma$.
Therefore the Coulomb effects have a much smaller energy
scale and the excited states do not participate
in mixing the ground states of the two potential wells.

At a nonzero temperature these Coulomb effects will compete
with the smearing by temperature so that
the width of the singularity at $\alpha=\pi$ is given at nonzero
temperature $T\ll\Delta$ by
$|\alpha-\pi| \sim \max(\gamma(N),T/\Delta)$.
Therefore, in order for Coulomb effects
to dominate the thermal
fluctuations, we must have $T\le\gamma\Delta$.

It is instructive to compare this picture with the case of
multi-channel tunnel S-S-S
junction (to distinguish from the results of \cite{GM} we should
remark that we consider the opposite to their assumption $\Delta<E_C$
limit). 
If we develop a similar theory for tunnel Josephson junctions,
we find that the potentials (\ref{potential}), (\ref{groundstates})
are both sinusoidal, and, therefore, the total potential
(\ref{potential}) has only one minimum (versus two in the
nearly ballistic system). In the tunnel S-S-S system
the current-phase relation $I(\alpha)$ has a smearing at
$\alpha=\pi$ due to the difference between the critical currents
of the two Josephson contacts. The Coulomb effects
compete with this smearing and in order to win, the
charging energy $E_C$ must be greater than the difference
of the critical currents. In the tunnel system the corresponding
splitting $\gamma$ is linear in $E_C$ while in the nearly
ballistic system it is exponentially small. Otherwise,
Coulomb oscillations in $I(\alpha)$ will appear similar in these
two cases.

To summarize this section, we observed that the Coulomb
effects in the one-channel S-S-S junction smears the singularity
in the Josephson current $I(\alpha)$ at the critical
value $\alpha=(2n+1)\pi$. This smearing depends
periodically on the potential of the grain with the period $2e/C$
 and is exponentially
small in the adiabatic parameter $E_C/\Delta\ll 1$. The smearing
is the result of mixing the two states in the potential minima
of the Josephson energy.

\section{Fluctuations of the Josephson current.}

In this section we compute the low-frequency spectrum
of the fluctuations of the Josephson current in our model.
We shall be interested in frequencies much less than the
oscillator energy scale $\sqrt{\Delta E_C}$,
thus we consider only transitions between the
eigenstates of the reduced ground-state Hamiltonian (\ref{Ham2}).
We also assume that the temperature is lower than
$\sqrt{\Delta E_C}$, then we may disregard the excited
oscillator states and the internal noise in the
contacts (discussed in \cite{MR1,AI,Gordei,MR3}). Obviously, under these
assumptions we can observe current fluctuations only
in the close vicinity of the resonance point $\alpha=\pm\pi$,
where the energies (\ref{energies}) of the two low-lying states
are close to each other.

We expect to observe two peaks in the noise spectrum ---
one at zero frequency (due to the thermal excitations above the
ground state), and the other at the transition frequency
$|E_1-E_2|$ (from the off-diagonal matrix elements of the
current operator). In this section we compute the intergal weights
of these peaks and postpone the discussion of their width
(determined by dissipative processes) until elsewhere.

Discuss first the zero-frequency peak. In our approximation
it is just  the thermal noise of a two-level system. In the
vicinity of the resonance point $\alpha=\pi$ we can linearize
the spectrum $V_{1,2}(\alpha)$ and make an approximation that
one of the two states carries the current $I(\alpha,N)$,
and the other $-I(\alpha,N)$. The spectral weight of the noise is then
given by a simple formula:
\begin{equation}
S_0(\alpha,N,T)\equiv \langle I^2 \rangle - \langle I \rangle^2
={I^2(\alpha,N) \over \cosh^2 {E_1-E_2\over 2T}}.
\end{equation}
Substituting $I(\alpha,N)$ and $E_{1,2}(\alpha,N)$
from the previous section,
we obtain the noise intensity near the resonance:
\begin{equation}
S_0(\alpha,N,T)={\Delta^2\over 2}\,
{ \left({\alpha-\pi\over2\sqrt2}\right)^2
\over  \left({\alpha-\pi\over2\sqrt2}\right)^2 + \gamma^2 (N)}
\cosh^{-2}\left({\Delta\over T}\sqrt{
\left({\alpha-\pi\over2\sqrt2}\right)^2 + \gamma^2 (N)}\right).
\label{S}
\end{equation}
For the effect of the Coulomb interaction to be observable, the
temperature must be smaller than the Coulomb gap: $T\le\gamma\Delta$.
At constant $T$ and $N$,
the noise decreases exponentially as $\alpha$ goes away
from its critical value $\alpha=\pi$, and at $\alpha=\pi$ the noise
is suppressed in the interval $|\alpha-\pi|<\gamma(N)$ (fig.~7).
Interplay between these two factors results in the strong dependence
of the peak value of the noise on the potential of the grain.
The peak value of the noise $\max_\alpha S(\alpha,N,T)$ is plotted
against $N$ in fig.~8. Most favorable is the case of identical
contacts, when $\gamma_1=\gamma_2=\gamma$ and, therefore,
$\gamma(N)=2\gamma\cos(\pi N)$. In this case,
when $\cos(\pi N)\ll T/\gamma\Delta$ (small
gap limit) the noise takes its maximal value $S\approx \Delta^2/2$.
In the opposite limit of large gap ($\cos(\pi N)\gg T/\gamma\Delta$)
the noise decreases exponentially: $S\approx \Delta^2 [{T\over\Delta
\gamma|\cos\pi N|} \exp (-4{\Delta\gamma|\cos\pi N| \over T})]$.
The noise has a sharp peak at the resonance point $\cos\pi N =0$,
where two levels on the grain with different electron numbers
have equal energies.


\begin{figure}
\centerline{\epsffile{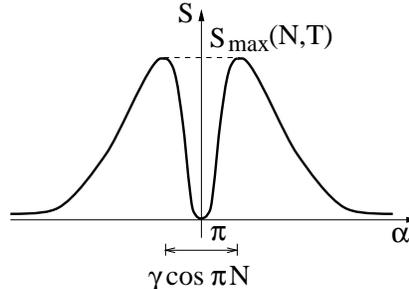}}
\caption{Zero frequency noise as a function of the phase $\alpha$.
It decays exponentially for $\alpha$ far from the resonance point
$\alpha=\pi$. At the very resonance point, the noise is suppressed,
because both of the two states carry nearly zero Josephson current.} 
\label{fig7}
\end{figure}

\begin{figure}
\centerline{\epsffile{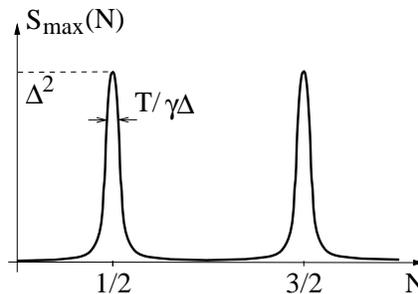}}
\caption{Maximal value of 
the noise versus the potential of the grain.
The period of the peaks corresponds to the period $2e$ of the
induced charge $Q=CV_g$. The width of the peaks depends on the
capacity of the grain.}
\label{fig8}
\end{figure}

Now turn to the noise peak at the interlevel frequency
$\omega=|E_1-E_2|$. Since now $\omega$ can be large compared to $T$,
one needs to discern between different kinds of frequency-dependent
correlation functions, which can be measured as a noise intensity in
 different experimental
situations \cite{lesovik}; here we mean by noise the Fourrier spectrum of
the time-symmetric current-current correlation function.
In our approximation of a two-level
system such a noise is temperature independent, and
its weight is determined purely by the off-diagonal matrix
element:
\begin{equation}
S_{\omega}={1\over 2}\Big|\langle 1|I|2\rangle \Big|^2.
\end{equation}
A straightforward computation for the Hamiltonian (\ref{Ham2})
and $I=2e(\partial H/\partial\alpha)$ gives
(in the vicinity of $\alpha=\pi$):
\begin{equation}
\langle 1|I|2\rangle={\Delta^2\gamma(N) \over\omega}
(\cos{\alpha\over4}+\sin{\alpha\over4})
\end{equation}
and
\begin{equation}
S_\omega(\alpha,N)=\Delta^2\left({\Delta\gamma(N)\over\omega}\right)^2
\cos^2{\alpha-\pi\over4}.
\end{equation}

This result contrasts the corresponding noise intensity
in the single quantum point contact (found in \cite{MR1,MR3,AI}).
In the single quantum point contact the correponding
noise intensity $S_\omega$ is temperature-dependent, because
that system has four possible states (or, alternatively,
two fermion levels). In the case of the double junction
the system has only two states differing by the phase
on the grain, and the quantum fluctuations $S_\omega$
become temperature-independent.

\section{Conclusions}

We have developed a theory of Coulomb oscillations of the Josephson current
and its noise power via the S-S-S system with nearly ballistic quantum point
contacts.  The period of Coulomb oscillations as function of the 
gate potential is  $V^0_g = 2e/C$.
These oscillations arise from the quasiclassical tunneling of the
superconducting phase on the grain and are, therefore, exponentially
small in $\sqrt{E_C/\Delta}$ at $E_C\ll\Delta$. 
In addition, we predict a crossover
from adiabatic to diabatic tunneling at the backscattering probability
$r_{ad}\sim \sqrt{E_C/\Delta}$. At backscattering below $r_{ad}$, the
amplitude $\varepsilon$ of the Coulomb oscillations is proportional to the 
square root of the
smallest (of the two contacts) reflection probability $\sqrt{r_{min}}$.
This constrasts the case of a normal double-contact system \cite{MF}
 where $\varepsilon$ is proportional to the product $\sqrt{r_1 r_2}$.

The average Josephson current-phase relation $I(\alpha)$ is shown to
be strongly non-sinusoidal and roughly similar to the one known for a
single nearly ballistic QPC, in the sense that it contains sharp "switching"
between positive and negative values of the current as the phase varies
via $\alpha = \pi$. The new feature of our system is that it is
possible to vary the width of the swithching region $\delta\alpha$
by the electric
gate potential $V_g$; in the case of equal reflection probabilities
$r_1=r_2$ this electric modulation is especially pronounced,
  $\delta\alpha \propto  |\cos(\pi CV_g/2e)|$.
The noise spectrum of the supercurrent is found to consist mainly of
two peaks: the "zero-frequency" peak due to rare thermal exitations of the
upper level of the system, and another one centered around 
the energy difference $\omega_{\alpha}$ between the two
levels. The widths of these peaks are determined by the inverse
 life-time $\tau$ of the two states of our TLS, which is due to electron-phonon
and electromagnetic couplings. Both these sources of level decay are
expected to be very weak in the system considered, but the corresponding
 quantitative analysis is postponed for the future studies, so we present 
here only the results for the {\it frequency-integrated} 
 (over those narrow intervals $\sim 1/\tau$) noise power.

The S-S-S device with almost ballistic contacts is a new type of 
a  system which may be used as a realization of an
artificial "spin 1/2" --- an elementary unit for quantum computations. 
In comparison with
usual Josephson systems with tunnel junctions  which 
were proposed for the use in adiabatic quantum computatons \cite{qaverin},
the advantage of our system  is that it may operate at considerably
higher values of the Josephson critical currents; moreover, the
current-phase characteristics of such a system is almost universal in the
sense that it is determined mainly by the microscopic parameters of the
SC materials and only weakly depends on the specifics of contact
fabrication.

We are grateful to K. A. Matveev, Yu. V. Nazarov and especially to
 G. B. Lesovik for many useful discussions. This research of M.V.F. 
was supported by the INTAS-RFBR grant \# 95-0302,
the collaboration grant \#  7SUP J048531
from the Swiss National Science Foundation and the DGA grant \# 94-1189.

\end{document}